\documentclass[prb,aps,twocolumn,floats,showpacs]{revtex4}
\usepackage{graphicx}

\newcommand{\nin}{\noindent}

\newcommand{\be}{\begin{equation}}
\newcommand{\ee}{\end{equation}}
\newcommand{\bea}{\begin{eqnarray}}
\newcommand{\eea}{\end{eqnarray}}
\newcommand{\lb}{\left[}
\newcommand{\rb}{\right]}
\newcommand{\lp}{\left(}
\newcommand{\rp}{\right)}

\newcommand{\la}{\left <}
\newcommand{\ra}{\right >}

\begin{document}
\title{L\'evy statistics and anomalous transport in quantum dot arrays}
\author{D.S.~Novikov$^{1,3}$}
\email{dima@alum.mit.edu}
\author{M.~Drndic$^{1,4}$}
\author{L.S.~Levitov$^1$}
\author{M.A.~Kastner$^1$}
\author{M.V.~Jarosz$^2$} 
\author{M.G.~Bawendi$^2$}
\affiliation{$^1$Department of Physics, $^2$Department of Chemistry,  
Center for Materials Sciences \& Engineering,  
%Massachusetts Institute of Technology, 
MIT, Cambridge, MA 02139\\
$^3$Department of Electrical Engineering and Department of Physics, 
Princeton University, Princeton, NJ 08540\\
$^4$Department of Physics, University of Pennsylvania, Philadelphia, PA 19104}

\date{\today}

\begin{abstract}
\nin
A novel model of transport is proposed to explain power law current
transients and memory phenomena
observed in partially ordered arrays of semiconducting nanocrystals. 
The model describes electron transport by a {stationary} L\'evy process 
of transmission events and thereby requires no time dependence of system
properties.
The waiting time distribution with a characteristic long tail gives 
rise to a nonstationary response in the presence of a voltage pulse. 
We report on noise measurements that agree well with the predicted 
non--Poissonian fluctuations in current, and discuss possible  
mechanisms leading to this behavior.

\end{abstract}

\maketitle

%%%%%%%%%%%%%%%%%%%%%%%%%%%%%%%%%%%%%%%%%%%%%%%%%%%%%%%%%%%%%%%%%%%%%%%%

%A variety of unusual transport phenomena are encountered in
%disordered electronic systems.
%Some of the transport mechanisms, 
%such as variable range hopping conductivity~\cite{MottDavis}, 
%are time independent, whether or not they involve electron-electron 
%interactions~\cite{Shklovskii}.
%However, when the many-body ground state is determined predominantly by 
%electron-electron interactions, such as in the Coulomb glass, 
%relaxation to the ground state can be slow;
%memory effects in conductivity have been attributed to 
%this relaxation~\cite{Ovadyahu}.
%Another situation in which the time dependence of the state of the system 
%leads to time dependent transport is when the charge carriers distribute 
%themselves among localized states as time progresses.  
%This leads to a power-law time decay of the current after excitation 
%with a light pulse, for example, 
%and is called dispersive transport~\cite{dispersion-transport}.

\section{Introduction}
\label{sec:intro}

\nin
Arrays of semiconductor nanocrystals~\cite{Bawendi} (quantum dot arrays, or QDAs)
are of great interest for both the fundamental solid state physics and applications. 
Self--assembled QDAs is one of the simplest examples 
of macroscopic complex systems built from the ``artificial atoms''
with pre--designed properties at the nanoscale. 
From the basic research perspective, these arrays are compelling 
since one can control the Hamiltonian by design. 
%Electronic transport is the basic tool to study nanocrystal arrays. 
In particular, they open new possibilities to create 
systems with desirable unconventional transport properties. 
Charge and spin transport in QDAs 
%Recent advances inlcude 
%improving these materials for their potential use in quantum electronics[Ref??]. 
%Observed coherent spin transfer in CdSe QDAs 
could lead to applications in spintronics and quantum computation. 
\cite{spintronics}

Despite the progress in synthesis and fabrication of nanocrystal arrays, 
the nature of electronic transport in them is still poorly understood.
%% unsettled.
%basic mechanisms of electronic transport in them are still not understood. 
%\mpar{*} 
%due to complexity of the system. % for both theorists and experimentalists.  
Like other problems involving long--range Coulomb interactions in disordered 
systems, this problem appears to be challenging.
Proposed theoretical models include mapping 
onto the problem of interface dynamics,\cite{Wingreen} 
%mapping of ordering and correlated hopping of charges
%on a triangular lattice 
onto a frustrated antiferromagnetic spin model
with long range interactions,\cite{Novikov01} 
as well as recent generalizations~\cite{ZhangShklovskii} 
of the variable range hopping scenario.\cite{conventional}

%in random resisor network~\cite{MA} due to 
%variable range hopping~\cite{MottDavis} in the presence % Efros--Shklovskii 
%of the Coulomb gap~\cite{EfrosShklovskii} 
%developed earlier for conventional semiconductors.
%The richness of the system brings together ideas and methods 
%from different domains of condensed matter physics, 
%as well as from optics and chemistry. 

In the present work we attempt to explain  
the recently observed\cite{qdots-transport,nicole}
transient power--law decay of current
\be \label{I-exp}
I(t) = I_0 \; t^{-\alpha} \, , \quad  0 <\alpha < 1   
\ee
as a response to a step in large bias voltage applied across the array.
The exponent $\alpha$ depends on temperature, 
dot size, capping layer, bias voltage and gate oxide thickness
in a systematic way.\cite{qdots-transport,nicole}
It is interesting that the observed $\alpha$ is less than one in all samples.
The condition $\alpha<1$ ensures that 
(\ref{I-exp}) is a true current from source to drain, 
rather than a displacement current, since
the net charge  
corresponding to (\ref{I-exp}) diverges with time,
$Q=\int I(t)dt \to \infty$.
%Already for observation times $\sim 10^3$~s, 
%the transported charge
%is orders of magnitude greater than that capacitively accumulated on the array.
%%DN as shown in Fig.~\ref{fig:transient}.
At the same time, remarkably, the transient transport %(\ref{I-exp}) 
also possesses {\it memory}. 
Namely, if the bias is turned off for $t_1 < t < t_2$, then
the current measured as a function of the time 
$\tilde t = t - t_2$ after reapplying the voltage
follows the dependence 
%% is of the form 
(\ref{I-exp}), albeit with a reduced amplitude $\tilde{I_0} < I_0$:
$I(t)=\tilde{I_0}(t - t_2)^{-\alpha}$.
%%DN 
%This is illustrated in Fig.~\ref{fig:transient}, 
%in which the transient for long times is recorded after that for the shorter time, 
%giving rise to a smaller amplitude.
%The amplitude $\tilde{I_0}$ is 
The amplitude is restored to its initial value, $\tilde{I_0} \to {I_0}$, 
by increasing the {\it off} interval $t_2-t_1$, by annealing at elevated temperature,
or by applying a reverse bias or band gap light between $t_1$ and $t_2$.
\cite{qdots-transport,nicole,ginger}

The behavior (\ref{I-exp}) is observed in partially ordered multi--layered 
arrays of II--VI semiconductor nanocrystals. %(e.g. CdSe, CdTe) 
%capped with $\sim$1\,nm coating.
Each nanocrystal is capped with $\sim$1\,nm coating, 
so that electrons must tunnel to move between neighboring sites.  
%Although these quantum dot arrays (QDAs) have short-range close-packed order, 
%they appear to have no long-range order. 
%In CdSe QDAs, since the zero bias conductance 
%is immeasurably small, transport properties have been studied 
%using strong applied fields~\cite{qdots-transport}. 
Although the transient response (\ref{I-exp}) has been 
recorded by a number of 
groups,\cite{qdots-transport,ginger,nicole,sionnest-science,sionnest}
its origin remains a mystery. 
In many systems the transient (\ref{I-exp}) comprises the dominant 
contribution to transport, while the ohmic conductivity is quite 
small.\cite{Marija-AFM} 
Understanding the nature of the time--dependent 
response (\ref{I-exp})
may shed light on the dynamics of carriers in such systems.

%Recently a new kind of disordered material has become available.  
%Progress in colloidal chemistry~\cite{Bawendi} 
%has made it possible to create novel solids, 
%composed of arrays of semiconducting nanocrystals. 
%Each nanocrystal has an organic coating, 
%so that electrons must tunnel to move between neighboring nanocrystals.  
%Although these quantum dot arrays (QDAs) have short-range close-packed order, 
%they appear to have no long-range order. 
%In CdSe QDAs, since the zero bias conductance 
%is immeasurably small, transport properties have been studied 
%using strong applied fields~\cite{qdots-transport}. 
%%As measurements \cite{qdots-transport,ginger,nicole,sionnest} point out,
%%the dc current in QDAs decays with time after 
%%a sudden application of a large voltage, 

It has been suggested earlier that the observed time--dependent 
current could be a result of time dependence of the state of the system. 
The latter could arise either because of charge flow jamming, due to
trapping of electrons blocking further charge injection 
from the contact,\cite{ginger} 
or because of the Coulomb glass behavior of the electrons distributed 
over QDAs.
\cite{qdots-transport}
However, it seems that such a scenario would require an unlikely 
fine--tuning as a function of time.
Namely, the system's properties would have to adjust 
in a coherent fashion over many hours to yield 
well--reproducible power laws in current, observed over at least 
five orders of magnitude in time in a broad variety of samples.

The purpose of this work is to suggest an alternative point of view on 
transport in QDAs, which does not require time--varying system's properties. 
We propose a model, based on the L\'evy statistics of waiting 
times between charge transmission events, 
in which the system remains {\it stationary} in statistical sense, 
but nonetheless exhibits a transient response. 
The model is corroborated by the measurements of the 
%% DN
%power law 
spectrum of time--dependent current fluctuations
in CdSe QDAs, and a good agreement %with the prediction of the model 
is demonstrated.

This paper is organized as follows.
We begin with introducing a phenomenological model of transport
which yields the response (\ref{I-exp}).
We consider manifestations of this transport mechanism 
in the noise spectrum, and report the results of noise measurements.
Finally, we
briefly discuss possible microscopic mechanisms that could be consistent with 
our transport model.

%%%%%%%%%%%%%%%%%%%%%%%%%%%%%%%%%%%%%%%%%%%%%%%%%%%%%%%%%%%%%%%%%%%%%%%%%%%%%%%%%%%
\section{Model of transport}
\label{sec:model}

\begin{figure}[t]
\centerline{
\begin{minipage}[t]{3.5in}
\vspace{0pt}
\centering
\includegraphics[width=3.5in,height=3in]{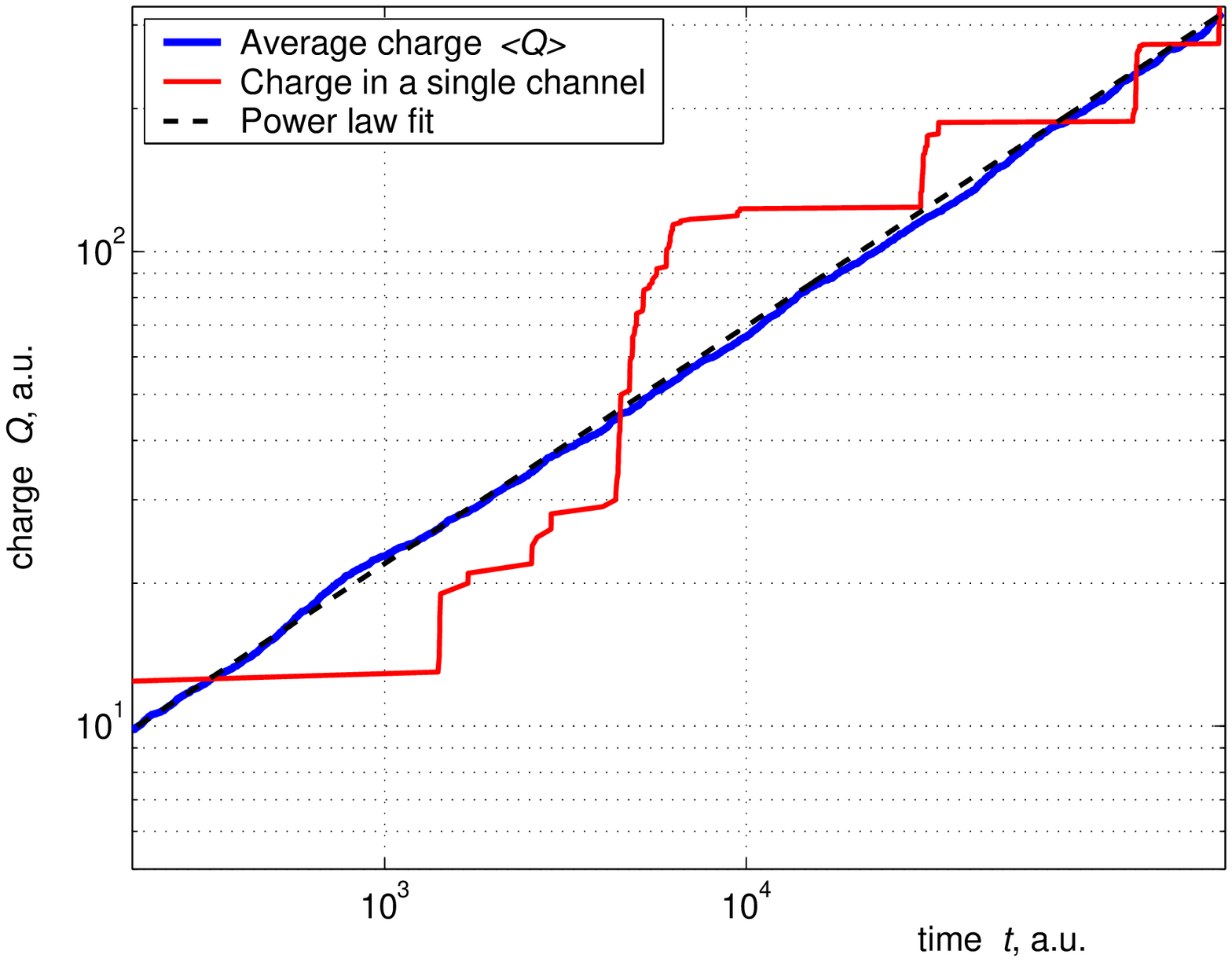}
\end{minipage}
\hspace{-1.85in}
\begin{minipage}[t]{1.8in}
\vspace{1.36in}
\centering 
\includegraphics[width=1.6in]{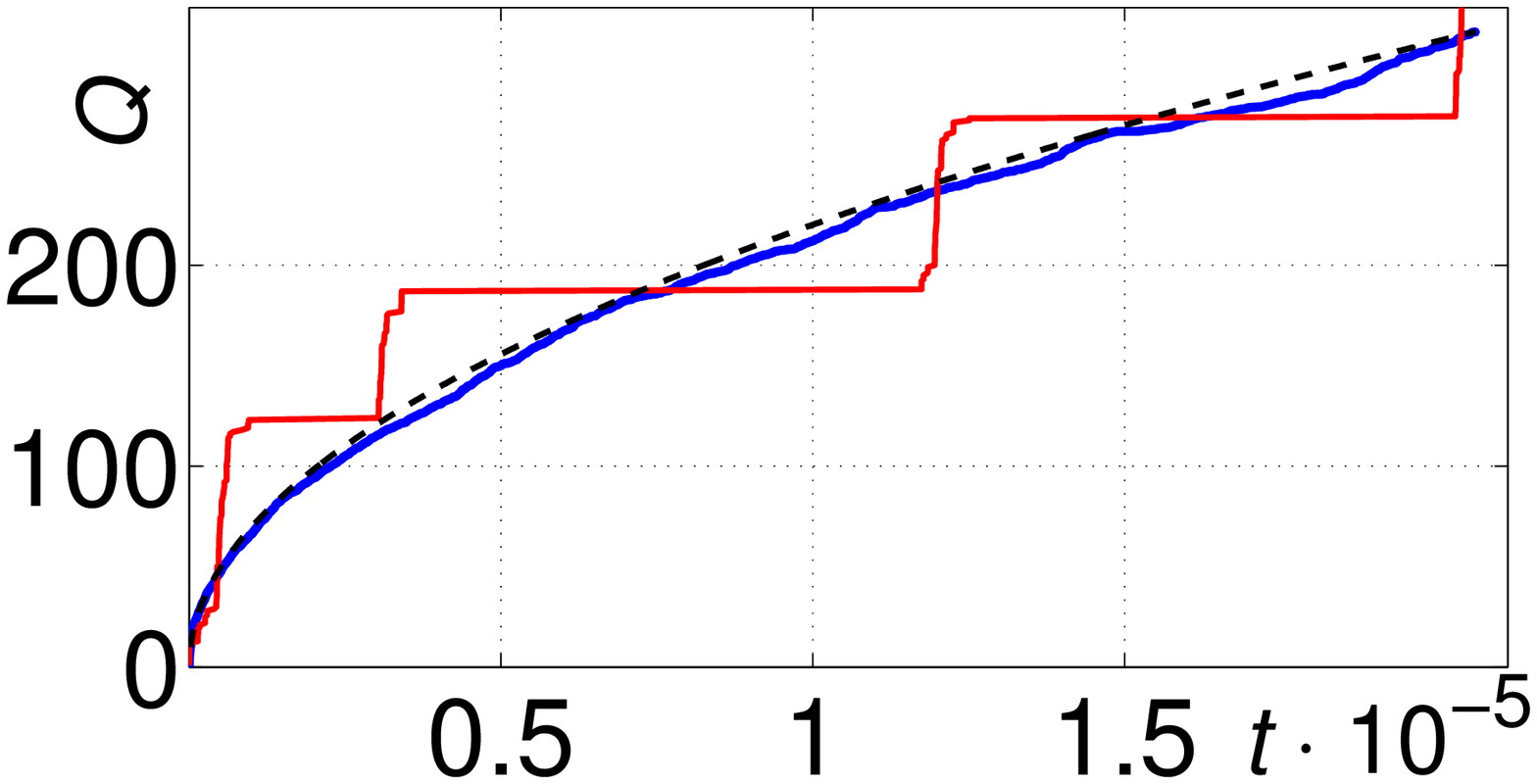}
\end{minipage}
\hspace{-1.9in}
\begin{minipage}[t]{1.8in}
\vspace{2.26in}
\centering 
\includegraphics[width=1.6in]{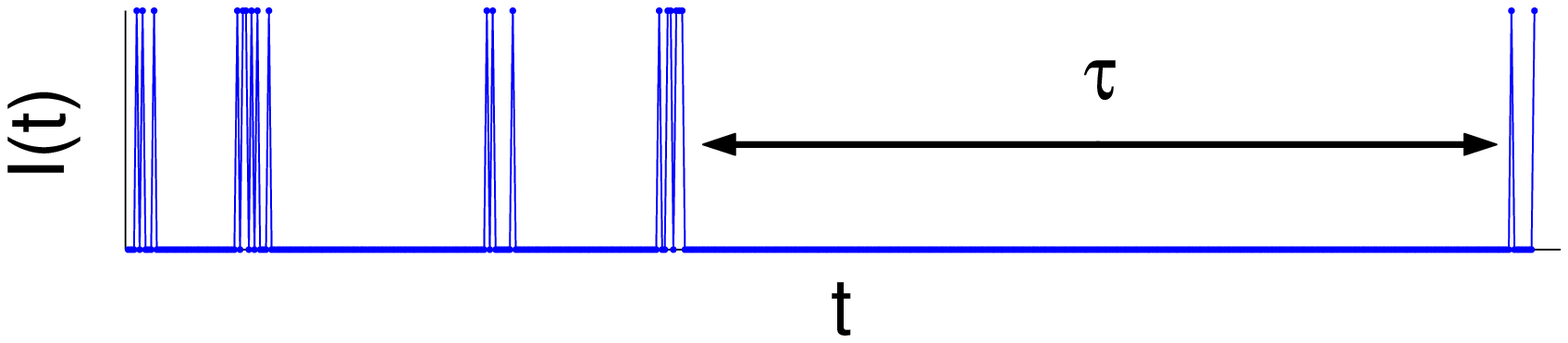}
\end{minipage}
}
\caption[]{\label{fig:ave-charge}
Time dependence of 
the net transmitted charge $Q(t)$ in a single channel (red line) and charge 
$\la Q(t)\ra$ averaged over $N=100$ channels (blue line)
simulated using WTD %waiting time distribution 
of the form (\ref{def-p}) with $\mu=0.5$ (double log scale).
Dashed line is a power law $Q\propto t^{\mu}$.
{\it Upper inset:} 
The same plot in the linear scale. 
The large charge noise in a single channel is due to 
the lack of self-averaging for a wide-tail WTD. 
{\it Lower inset:}  Current in a single channel with a wide distribution of 
waiting times (schematic). Short packets of
%($\tau_0$) 
current pulses are separated by very long waiting times $\tau\gg\tau_0$.

%% It does not at all resemble the average $\la Q\ra$.
%over $N$ channels.
}    
\end{figure}

\nin
The main idea of our approach is that current (\ref{I-exp}) can arise 
as a result of a {\it stationary} stochastic process.
Our model involves $N \gg 1$ identical 
independent conducting channels arranged in parallel. 
(This accounts for the typical sample's 
aspect ratio $\sim 10^3\,\mu$m\,:\,$1\,\mu$m.)
Each channel is almost always closed, 
and opens up at random for a short interval $\tau_0$
to conduct a current pulse that corresponds to a unit transmitted charge,
as schematically shown in the lower inset of Fig.~\ref{fig:ave-charge}. 
We further assume that the intervals between subsequent transmissions 
are uncorrelated, making the process completely characterized by the 
{\it waiting time distribution} (WTD) 
%% $p(\tau)$
of intervals between successive pulses. 

In particular, we will be interested in WTD $p(\tau)$
with a broad tail at long times.
In order to model the power law decay of the current transient,
here we consider a special form of WTD with a long tail of the L\'evy type:
\be \label{def-p}
p(\tau\gg \tau_0) \simeq \frac{a}{\tau^{1+\mu}} \, , 
\quad 0 < \mu < 1 \, ,
\ee
with $\tau_0$ a miscroscopic time scale.
Note that {\it all} moments of $p(\tau)$ diverge.
The behavior of the WTD at short times, $p(\tau \sim \tau_0)$, is not of interest, 
since it does not affect the long time dynamics.

%%%%%%%%%%%%%%%%%%%%%%%%%%%%%%%%%%%%%%%%%%%%%%%%%%%%%%%%%%%%%%%%
%%%%%%%%%%%%%%%%%%%%%%%%%%%%%%%%%%%%%%%%%%%%%%%%%%%%%%%%%%%%%%%%

As shown in Appendix~\ref{sec:noise-calc}, 
WTD of the form (\ref{def-p}) indeed yields the power law decay 
(\ref{I-exp}) for the mean value of the current in a single channel
with $\alpha = 1-\mu$ and $I_0=\mu \sin \pi \mu/\pi a$. 
Qualitatively, the decrease in current with time can be understood as follows.
The mean value of the waiting time for the process with WTD (\ref{def-p}) is 
{\it infinite}.
Thus, for a stochastic process governed by the WTD (\ref{def-p}) 
which started infinitely early in the past,
the observed value of the current would be zero.
Turning the bias on at $t=0$ sets the clock for the process. %(\ref{def-p}). 
In this case, for the measurement interval $t$, only the waiting times 
$\tau \le t$ can occur,
as illustrated by the simulation shown in 
Fig.~\ref{fig:ave-charge} (note the double log scale).
Observing the current over a larger time period effectively increases the 
chances for a channel to be closed for a longer time interval, which results
in the
decay in the average current, the latter approaching zero at $t\to\infty$. 
We note that in this transport model the system's parameters 
characterizing the distribution (\ref{def-p}) are {\it time independent}. 
Hence the process %(\ref{def-p}) 
is {\it stationary}, i.e. each charge transmission event occurs after
a delay time $\tau$ described by the distribution
$p(\tau)$ independent of the total time $t$ passed after the beginning
of the measurement.

%Continuous time random walks with long power law tails as in (\ref{def-p})
%(the so called L\'evy processes \cite{Levy-flights}) arise in various 
%contexts \cite{Bouchaud}, from 
%dispersive transport of photocurrent in amorphous semiconductors 
%\cite{dispersion-transport}
%to the stock market fluctuations \cite{finance}. 
%The main feature of such processes is that, due to divergent moments,
%they violate the central limit theorem. 

Continuous time random walks with the L\'evy WTDs 
often arise in the systems characterized by wide distributions of time 
scales.\cite{Bouchaud}
In the semiconductor physics the L\'evy processes 
have been extensively studied in the 
context of the {\it dispersive transport}, e.g. in amorphous semiconductors.
\cite{dispersive-transport}
A simple example is a system of electrons moving between charge traps.
Its dynamics depends on energy $\epsilon$ via an activation exponential, 
$\tau = \tau_0\, e^{\beta\epsilon}$, where $\beta$ is the inverse temperature. 
For the distribution of the energies $\epsilon$ described by the density of states 
of exponential form, $\nu(\epsilon)=\nu_0 e^{-b\epsilon}$, one obtains
$p(\tau)$  of the power law form (\ref{def-p}) with exponent
$\mu=b/\beta$ and $a=\nu_0 \,\tau_0^{\mu}/\beta$.

The probability distribution (\ref{def-p}) leads to an unusual behavior 
which is the subject of the theory of L\'evy flights. \cite{Levy-flights}
The main characteristic of the L\'evy statistics\cite{Levy-flights,Bouchaud} is the 
violation of the central limit theorem.
To illustrate this unconventional behavior, let us recall what happens in 
a Poissonian channel characterized by the finite mean waiting time $\bar\tau$.
The mean value of the transmitted charge $Q$  
% between $t=0$ and $t=T$ 
%is $\la Q(T)\ra = T/\bar \tau$, 
grows linearly with time, $\la Q\ra = t/\bar \tau$,
corresponding to a constant current.
The variance of the charge is proportional to the mean,
$\la\!\la Q^2\ra\!\ra = \frac12 \la Q\ra$, in other words
the relative charge fluctuation decreases,
$\la\!\la Q^2\ra\!\ra^{1/2}/\la Q\ra \propto t^{-1/2}$,
in accord with the central limit theorem. 
In contrast, in the case of the distribution
%of the L\'evy form 
(\ref{def-p}),
%charge cumulants at $t\gg \tau_0$ 
%are anomalously large: $\la\!\la Q^n\ra\!\ra \sim \la Q\ra^n$,
the mean transmitted 
charge increases {\it sublinearly} as $\la Q\ra \sim t^{\mu}$, whereas 
its variance is proportional to the {\it square} of the mean,
$ \la\!\la Q^2\ra\!\ra \propto \la Q\ra^2$ 
(see Appendix~\ref{sec:noise-calc}). 
Since relative charge fluctuation does not decrease with time,
%$\Delta Q(T) = \lp \la Q^2\ra-\la Q\ra^2\rp^{1/2} \propto \la Q(T)\ra$.
transport in a single channel 
%% with the WTD (\ref{def-p}) 
is dominated by large fluctuations of waiting times
(see Fig.~\ref{fig:ave-charge}).

Although in our model any given channel lacks self--averaging,
the charge summed over $N\gg 1$ independent parallel channels averages to a 
smooth power law, Fig.~\ref{fig:ave-charge}, with fluctuations 
reduced by a factor of $N^{-1/2}$.
%Fluctuations in the total current through a system with $N$ parallel channels
%are reduced by a factor of $N^{-1/2}$, as shown in Fig.~\ref{fig:ave-charge}
%for $N=100$. 
For the typical sample geometry used in our experiments, consisting of 
{\it ca.} 50 layers, 
each layer of $1.6\cdot 10^5$ dots wide and 200 dots across, one expects 
large effective $N$ and small current fluctuations, 
%$N\sim10^2-10^4$,
%and the fluctuations in current to be small,
as in Refs.~\onlinecite{qdots-transport,nicole}.
%%DN
%Fig.~\ref{fig:transient}.

Also, as a check of robustness of this scenario with respect to 
spatially varying system properties, we considered parallel 
non--identical channels, characterized by non--equal values of $\mu$. 
We found that the average current obtained from such a model 
is approximately described  by a power law of the form (\ref{I-exp}).
We performed a simulation with $N=100$ channels, 
with the exponents of different channels
drawn from a flat distribution, $0.45<\mu<0.55$.
In this case, the transmitted charge time dependence
was found to be numerically very close to that
given in Fig.~\ref{fig:ave-charge} with $\mu=0.5$.
%%that in Fig.~\ref{fig:transient} for $\mu=0.5$.

%%%%%%%%%%%%%%%%%%%%%%%%%%%%%%%%%%%%%%%%%%%%%%%%%%%%%%%%%%%%%%%%%%%%%%%%%%%
\section{Memory Effects}
\label{sec:memory}

\nin
Memory effects originate in our transport model 
in the manner analogous to aging in the L\'evy systems.\cite{barkai'03}
Because of large typical waiting times $\tau\gg \tau_0$,
any given channel is most likely found in a non--conducting state
when the voltage is turned off at $t = t_1$.
In addition, we assume that,
due to gradual time variation of channel parameters,
taken to be very slow in this discussion, 
the channel state is likely to remain 
unchanged by the time the voltage is turned back on at $t=t_2 > t_1$.
In this case the channel conducts current as if 
the voltage has been on all the time.
However, due to the aforementioned time variation of system parameters,
there is a chance that the channel changes 
its state (resets) while the voltage is turned off during 
$t_1<t<t_2$.
This reset probability 
$w_{12}\equiv w(t_2 - t_1)$, as a function of the off time
$t_2 - t_1$, 
is growing monotonically:
$w(\tau_0) \approx 0$, $w(\infty) = 1$.
As a simple model of this behavior, one can consider a Poisson process,
\[
w(\tau)=1-e^{-\gamma t}
\]
with the rate parameter $\gamma$ characterizing the reset probability.

The current at $t = t_2$
as a function of the {\it shifted time} $\tilde t  = t - t_2$,
obtained by averaging over $N\gg 1$ channels, is given by
\be \label{I-mem}
I(\,\tilde t\,) = (1-w_{12})I_0\, (\tilde t + t_2)^{-\alpha} 
\, + \, w_{12}I_0\, \tilde t^{-\alpha} \, .
\ee
Here we assume the reset of different channels to be independent 
and uncorrelated.
The function $I(\tilde t)$ has a singular part at $\tilde t \approx 0$ 
(the second term of Eq.(\ref{I-mem}))
with the amplitude 
$\tilde I_0 = w_{12}I_0$ reduced compared to $I_0$
by the reset probability $w_{12}<1$. 
Thus at $\tilde t\ll t_2-t_1$
%% For $t_2\gg \tau_0$, 
the first (regular) term in Eq.~(\ref{I-mem}) is negligible 
compared to the second term. The current (\ref{I-mem})
is dominated by the latter, resulting in
%% an apparent 
suppression of the measured transient current part which is singular
at $t\approx t_2$.

We note that the described reset process, 
while leading to suppression of the singular part of the current, 
is accompanied by an
overall enhancement of the total current (\ref{I-mem}),
as compared to the current (\ref{I-exp})
%as compared to the current (1) 
at time $t$ in the absense of resetting. 
This prediction indeed agrees with our observations.
We have verified that the reset probability 
$w_{12}=\tilde I_0/I_0$ is indeed a monotonic function
of the time interval when the voltage is turned off. 
For waiting times from $10\,{\rm s}$ to $10^{4}\,{\rm s}$ 
in between $100\,{\rm s}$ long transients, we measure 
$0.65 < w_{12} < 0.85$; 
%in the range 0.65 to 0.85. 
$w_{12}\to 1$ when applying a reverse 
bias, exposing the dots to the band gap light %\cite{nicole},
or waiting for longer times.

%%%%%%%%%%%%%%%%%%%%%%%%%%%%%%%%%%%%%%%%%%%%%%%%%%%%%%%%%%%%%%%%%%%%%%%%%%%%%%%
\section{Noise frequency spectrum}
\label{sec:noise-th}

\nin
%% Although t
The model described above, which is consistent with previously 
reported transport measurements, 
%% it needs to 
can be independently verified with the help of noise measurements.
Here we consider the statistics 
of current {fluctuations} and 
formulate a prediction of the model based on
the L\'evy process (\ref{def-p}).
%% that has to do with the statistics 
%% of current {\it fluctuations}, rather than with current itself. 

The unconventional fluctuations exhibited by the L\'evy process,
discussed in Section~\ref{sec:model}, manifest themselves
in noise as follows. 
Consider the time-dependent current
in a single channel, described in Section~\ref{sec:model},
recorded during a long time interval %$0< t < T$, 
$T\gg \tau_0$:
\be \label{def-I1}
I(t) = \sum_{n=1, 2, ...}
%\sum_{i=1}^{\infty} 
\delta(t-t_n) \, , \quad 0 < t < T \, . 
%\theta(T-t)
%\,e^{-\lambda t} \, , \quad \lambda = T^{-1} \, . 
\ee
The intervals $\tau_n = t_n - t_{n-1}$, $n=1,2, ...$, $t_0 \equiv 0$,
are independent random variables distributed according to the WTD 
of the form (\ref{def-p}).
The fluctuations of current 
%% $\la\la I_{-\omega}I_{\omega}\ra\ra$ 
are defined in terms of the Fourier harmonics
\be \label{def-I1-omega}
I_{\omega}=\int_0^T e^{i\omega t}  I(t)dt\, .
\ee
Here we consider the {noise power spectrum}
\be \label{def-S}
S(\omega)=\la\la I_{-\omega}I_{\omega}\ra\ra 
= \la I_{-\omega}I_{\omega}\ra  - \la I_{-\omega}\ra \la I_{\omega}\ra  \, .
\ee
In Appendix~\ref{sec:noise-calc} we show that 
%It can be shown that 
the distribution (\ref{def-p}) leads to the non--Poissonian spectrum %in the current,
\be \label{noise}
S(\omega)\propto
%%=\la\la I_{-\omega}I_{\omega}\ra\ra \propto 
\cases{ 
T^{2\mu} \, ,
& $\omega T \ll 1$ , 
\cr
T^{\mu} \omega^{-\mu} \, , 
 & $\omega T \gg 1$ .
} 
\ee
Here the low frequency part of (\ref{noise}) 
with $\omega T \ll 1$ corresponds to the
fluctuation $\la\!\la Q^2\ra\!\ra$ 
of the net transmitted charge. Due to the relation 
$\la\!\la Q^2\ra\!\ra \propto \la Q(T)\ra^2$ [Eq.~(\ref{disp-Q})]
between the first two moments of the L\'evy process, %discussed above,
which violates the central limit theorem, the quantity $S(\omega T \ll 1)$
is proportional to the square of the mean transmitted charge $Q(T)\propto T^{\mu}$. 

Experimentally, however,
it is more convenient to deal with $S(\omega)$ at finite frequency
$\omega T \gg 1$. Eq.~(\ref{noise}) predicts a characteristic 
power law spectrum for this quantity.
We note that for $\omega T \gg 1$, 
the L\'evy process (\ref{def-p}) yields \emph{identical} 
power laws for the noise spectrum (\ref{noise})  
and for the average current, $\la I_{\omega} \ra \sim \omega^{-\mu}$.
Moreover, the relationship 
\be \label{prediction}
\la\la I_{-\omega}I_{\omega}\ra\ra \propto \la I_{\omega}\ra \propto \omega^{-\mu} 
\, , \quad \omega T \gg 1 \, ,
\ee
is robust with respect to averaging over $N$ independent channels,
since for such averaging the central limit theorem holds. 
An experimental test of the proportionality relationship (\ref{prediction}) 
between the 
frequency spectra of current and noise will be discussed below.

%%%%%%%%%%%%%%%%%%%%%%%%%%%%%%%%%%%%%%%%%%%%%%%%%%%%%%%%%%%%%%%%%%%%%%%%%%%%
\section{Noise measurements}
\label{sec:noise-exp}

\nin
%{\it Noise measurements.---}
Here we briefly describe the experiments performed to obtain
the data on noise 
frequency spectrum in QDAs.
The QDAs were produced as described 
in Ref.\onlinecite{qdots-transport} by self-assembly of nearly 
identical CdSe nanocrystals, 3 nm in diameter, capped with trioctylphosphine 
oxide, an organic molecule about 1\,nm long.  A film of about 200 nm thick 
of the nanocrystals was deposited on oxidized, degenerately doped Si wafers 
with oxide thickness $\approx$ 200\,nm. 
%%DN (see the inset of Fig.~\ref{fig:transient}).  
%%DN added:
The experimental setup was similar to that utilized 
in Ref.~\onlinecite{qdots-transport}.
Gold electrodes, fabricated on the surface before deposition of the QDA, 
consist of bars 800$\mu$m long with separation of 2$\mu$m.  
The sample was annealed at 300\,C 
in vacuum inside the cryostat prior to the electrical 
measurements.  Annealing reduces the distance between the 
nanocrystals and enhances electron tunneling.\cite{qdots-transport}

%%%%%%%%%%%%%%%%%%%%%%%%%%%%%%%%%%%%%%%%%%%%%%%%%%%%%%%%%%%%%%%%%%%%
\begin{figure}[t]
\includegraphics[width=3.5in]{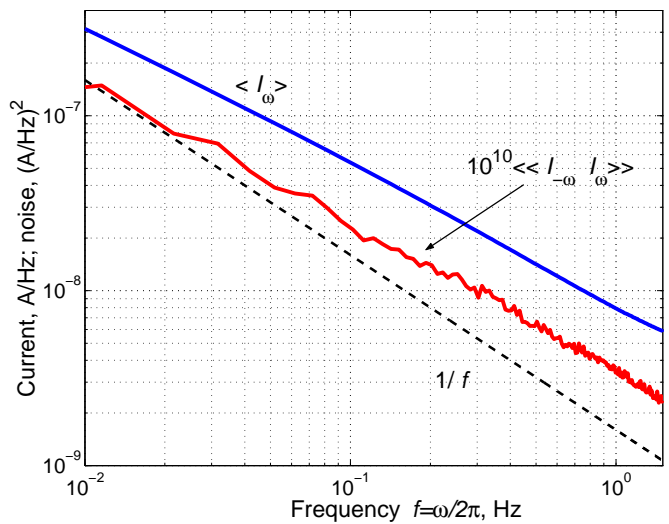}
\caption[]{\label{fig:noise}
Measured mean current  
and noise spectrum, with averaging 
performed over 50 current transients on the same sample.
The current Fourier harmonic mean $\la I_{\omega}\ra$ [Eq.~(\ref{def-I1-omega})] 
and variance $\la\la I_{-\omega}I_{\omega}\ra\ra$ [Eq.~(\ref{def-S})]
are shown.
Both the current and the noise are described by power law 
$\omega^{-\mu}$ with the same $\mu\approx 0.72$.
The $1/f$ dependence is shown for comparison (see text).
}
\end{figure}
%%%%%%%%%%%%%%%%%%%%%%%%%%%%%%%%%%%%%%%%%%%%%%%%%%%%%%%%%%%%%%%%%%

To measure the noise, we have recorded 200 current transients each $t=100$~s long.
Measurements have been made on a single sample continuously stored in 
vacuum, inside of a vacuum cryostat in the dark at $77$~K.
Each current transient was recorded for 100~s with a negative bias 
of $-90$~V. These periods of negative bias were separated from each other 
by a sequence of zero bias for 10~s, reverse pulse of $+90$~V 
for 100~s, and zero bias for 10~s, to eliminate the memory 
effects.\cite{qdots-transport}
We checked that current fluctuations for a substrate without the QDA were
several orders of magnitude smaller than with the QDA.
%We have measured the current and noise for a substrate without the QDA 
%and find that the typical current fluctuations 
%are several orders of magnitude smaller than those with the QDA. 

%An error in the average current 
%% $\la I_{\omega}\ra$ 
%can yield an error $\propto \la I_{\omega}\ra^2 \propto \omega^{-2\mu}$ 
%in the noise spectrum.
 
At the beginning of our measurement the current transients were 
changing from one to the next because of the memory effect described above.
Since an error in the average current can yield an error
of order
$\la I_{\omega}\ra^2 \propto \omega^{-2\mu}$ which may affect the measured
noise spectrum power law, we discarded the first 150 transients.  
The noise (Fig.~\ref{fig:noise}) 
was then deduced from the remaining 50 transients. 
%whose average current is nearly unchanged from one to the next.
To further compensate for residual memory effects, each transient was 
%normalized 
multiplied by a factor $\approx 1$ 
to have the same net integrated charge. 
This eliminated the zero frequency contribution to the noise.

Figure~\ref{fig:noise} shows the measured noise spectrum and 
the average current measured simultaneously. 
Both quantities have a power law behavior with 
nearly identical exponent values $\mu\approx 0.72$
for $\omega t\gg 1$, $t=100$\,s.
With relative $\sim3-10\%$ deviations of the noise from the $\omega^{-\mu}$ 
law, the measured noise spectrum is clearly distinct 
from the $1/f$ noise typically found at low frequencies.
For comparison, in Fig.~\ref{fig:noise} we draw the $1/f$ dependence,
offset so that it coincides with the noise data at the lowest frequency. 
The discrepancy with the measured noise 
at the highest frequency by more than a factor of two indicates
that the observations are not explained by the $1/f$ noise model.
%% it then differs from the measured noise by more than a factor of two. 
%The latter is drawn to coincide with the noise at low frequencies. 
%Both spectra differ by more than a factor of 2 at $f > 1\,$Hz.
%Curves coinciding at low frequencies differ by more than a factor of two 
%at the highest frequency.

One may question whether the observed colored noise,
instead of being a consequence of the L\'evy process, 
could result from interplay of the
intrinsic $1/f$ noise and the time-dependent current decaying 
according to (\ref{I-exp}).
%rather than a consequence of the L\'evy process. 
In this case the fluctuations
would be proportional to the current itself:
\be
I(t)=\la I(t)\ra(1+s(t))
\,,\quad
\la s(t)s(t')\ra \equiv \Gamma(t-t') \, ,
\ee
where 
%$\la s\ra=0$ and 
%% $\la s(t)s(t')\ra \equiv \Gamma(t-t')$, 
$\Gamma_{\omega}\propto \omega^{-\eta}$,
$\eta\simeq 1$ for the $1/f$ noise.
This would yield the current fluctuation spectrum of the form
%% dispersion of the current 
%
\be
\la\!\la I_{-\omega}I_{\omega}\ra\!\ra = 
\int |\la I_{\omega'}\ra|^2 \Gamma_{\omega'-\omega} 
\frac{d\omega'}{2\pi} \, .
\ee
When $\mu > 1/2$, the integral is dominated by the $(\omega')^{-2\mu}$ 
singularity, which effectively sets $\omega'=0$, giving rise to 
the $1/\omega$ behavior. 
Physically this happens because the current (\ref{I-exp}) for 
$\alpha < 1/2$ decays slowly enough so that all the harmonics of the $1/f$
noise have time to fully play out. 

Conversely, in a system with $\mu<1/2$, 
the noise $\propto \omega^{-2\mu}$ may be indistinguishable from the errors 
in determining average current described above.
For the consistency check of our model it is important
that for our sample the observed current transient power law exponent fulfills
$\mu>0.5$. The observation of the $\omega^{-\mu}$ noise 
indicates that transport is not dominated by the intrinsic 
$1/f$ noise. Instead, we conclude that
the noise measurement 
agrees with the proposed transport model based on
L\'evy statistics of transmission events.

% From the data file:
%
% measuring 2 um gap after a long time
% 10^8 gain on current amp, reanealed at 300c,
% 100s steps, voltage is 1.8V (1 V = 50v source drain)
%
% Ltrans2.1 open on: Wed Aug  7 16:40:35 2002

%Zero frequency behavior in Fig.~\ref{fig:noise} corresponding to noise in 
%the net charge is dominated by the residual memory effect, due to which 
%each successive transient has on average a slightly smaller amplitude $I_0$.
%This together with the fast initial capacitive charge accumulation 
%results in each successive transient having a different net charge 
%$I_{\omega = 0} = \la Q(T)\ra$. 
%Finite frequency measurements are insensitive to these effects and
%result in a power law with $\mu\approx 0.7$ for our sample.

\section{Discussion}
\label{sec:discussion}

%{\it Discussion.---}
%Let us consider possible microscopic origins of the 
%L\'evy process governing conductivity of the dot arrays.
%We turn to the standard transport mechanism in disordered semiconductors, 
%hopping conductivity \cite{MottDavis,Shklovskii}.
%There, phonon assisted hops between impurity sites 
%yield a random resistor network of Miller and Abrahams
%\cite{MA}. The latter results in a time independent and self averaging 
%resistivity of the sample.
%Such a variable range hopping resistivity 
%has an exponential temperature dependence \cite{MottDavis}, 
%that is further enhanced by the Coulomb gap effects \cite{Shklovskii}. 

%In the dot arrays ohmic conductivity 
%%%(possibly due to phonon-assisted processes) 
%is negligible \cite{qdots-transport,nicole,Marija-AFM}.
%%%Moreover, no exponential temperature dependence in transport was observed. 
%Variable range hopping is absent due to the geometry of the arrays,
%where hops mainly occur between nearest neighbors.
%Extremely slow diffusion in the ohmic regime ($D\sim (0.1\mu$m$)^2$/s) 
%\cite{Marija-AFM} points at a possible phonon bottleneck
%for such hops. 
%%%that there is not enough phonons to accommodate for such hops. 

\nin
%What is the effective number of channels?
Based on the noise measurements we can estimate the effective number $N$ of 
conduction channels introduced in Section~\ref{sec:model}.
Since both the measured current $\la I_{\omega}\ra$ and noise $S(\omega)$
are proportional to $N$, the noise-to-current 
ratio $r(\omega) \equiv S(\omega)/\la I_{\omega}\ra$ 
is a characteristic of a single channel.
%is the same as that for the single channel. 
The model calculation for the latter [Appendix~\ref{sec:noise-calc}] shows 
that at large frequency $\omega t \gg 1$ the noise-to-current ratio is
frequency independent and proportional to the net charge $Q(t)$ 
transmitted through the channel during the time $t$ of measurement, 
%$S/\la I_{\omega}\ra \sim Q(t) \equiv I_{\omega=0}$  %$ = A_{\mu} T^{\mu}$ 
$r(\omega t\gg 1) \sim Q(t)$  %$ = A_{\mu} T^{\mu}$ 
[Eq.~(\ref{r})]. 
From Fig.~\ref{fig:noise} we find that the measured $r(\omega)$ 
is indeed frequency independent for $\omega t\gg 1$, and is of the order 
$r \approx 10^{-10}\, {\rm A}/{\rm Hz}$.
%$S/\la I_{\omega}\ra \approx 10^{-10}\, {\rm A}\cdot {\rm s}$.
[Averaging over 50 transients does not affect $r(\omega)$.]
The effective number of channels is then estimated as the ratio 
of the measured net transmitted charge %through the system 
to that in a single channel,
$N \sim \la I_{\omega=0}\ra / r \sim 10^4$. This large number of independent
channels is consistent with the sample geometry 
(aspect ratio $\sim 10^3$ and {\it ca.} 50 layers of dots).

How can the long waiting times with a distribution of 
L\'evy form (\ref{def-p}) arise microscopically? 
While presently there is no fully satisfactory answer to this question,
one can make several observations. First,
to rationalize a wide distribution of time scales 
such as (\ref{def-p}), we suppose that the charge hops 
between neighboring dots are strongly constrained.
The simplest constraint to imagine 
is the lack of energy relaxation 
(possibly due to a small number of available phonon states) that arises 
if the on--site energies
of electrons on different dots are widely distributed. 
This is consistent with the absence of ohmic contribution to the conductivity
in our QDA.
%One could imagine that the immeasurably small zero bias conductivity in our QDAs
%could mean not having enough phonons to relax energy for charge hops.
%points at a possible phonon bottleneck in energy relaxation for the hops of 
%carriers between adjacent nanocrystals. One can speculate that such hops 
%responsible for the observed current transients 
The energy relaxation constraint then allows charge 
hops only between the {aligned} energy levels of the dots.

Next, the WTD (\ref{def-p}) with a long tail can be explained if
the energy levels strongly fluctuate in time,
with $\mu=\frac12$ corresponding to the Gaussian diffusion in energy. 
One can think of at least two reasons for the level fluctuations. 
First, the voltage bias energy $\sim 0.1$\, eV dissipated per hop 
may provide the necessary energy reservoir. % for the fluctuations.
Second, current--induced fluctuations in the electrostatic environment
in the absence of screening may result in a random time--dependent chemical
potential for each dot. 
%
%
%The WTD (\ref{def-p}) can arise in this setting even without level diffusion.
%
%With dot energy levels strongly fluctuating, 
%the waiting time for the electron to hop 
%between the adjacent dots is the time between 
%successive alignments of their levels.
%The level diffusion in energy itself can be governed by a continuous 
%time random walk.
%If the latter is described by the WTD of the type 
%(\ref{def-p}), $\phi(\tau) \sim 1/\tau^{1+\nu}$, and the distribution of 
%hops in energy is $\psi(\epsilon) \sim 1/\epsilon^{1+\kappa}$, 
%one obtains
%$\mu = \nu(\kappa -1)/\kappa$ for the nearest neighbor hopping WTD 
%(\ref{def-p}). 
%The Gaussian diffusion in energy yields $\mu=1/2$, as conjectured 
%in Refs.~\cite{shimizu,barkai}
%to explain the L\'evy statistics of a single dot fluorescence intermittency
%%observed in \cite{shimizu,nesbitt,dahan,bronkmann}.
%Finally, the WTD for the ``channel'' (a chain of sites across the sample) 
%has the same power law tail as the WTD for 
%the nearest neighbor hops in the absence of Hubbard on--site correlations. The 
%latter are believed to be unimportant since in \cite{qdots-transport,nicole} 
%the estimated electron density per dot is small, $\sim 10^{-2} - 10^{-1}$. 
%
%
%DN new reason!!
%Another microscopic reason for the process (\ref{def-p}) could be 
In particular, 
misalignment of the energy levels can arise due to the Coulomb field
of an electron trapped 
%due to a possibility of escaping of electron from the quantum dot to the 
%charge trap located 
in the vicinity, {\it e.g.} in the coatings. 
%Then the ``channel'' passing through this particular nanocrystal could 
%be closed if the Coulomb field of the trapped electron substantially
%shifts the dot's energy levels. 
The ``conduction channel'' then opens up when the trapped electron
escapes. Due to large applied bias, 
filling of the traps can happen much faster 
than escaping from them.
With escape times exponentially dependent on trap parameters, the 
distribution (\ref{def-p}) follows naturally.\cite{dispersive-transport}

We note that this picture differs somewhat from the canonical 
dispersive transport mechanism,~\cite{dispersive-transport}
in which a constant supply of carriers makes
the current grow with time.\cite{street} %at a constant bias.
The growth of current occurs 
due to the increasing number of
trapped electrons levelling the potential landscape,
thereby enhancing conductivity.
%This is explained by gradually filling in of the traps thus smoothing the 
%potential for the carriers, effectively increasing the sample's
%conductivity. 
Contrarily, in the proposed picture the presence of traps 
regulates the dynamics of conducting channels.

%The suggested transport explain 
We also note that the L\'evy statistics was recently
observed in fluorescence intermittency of {\it individual} nanocrystals.\cite{fluor}
%\cite{shimizu,nesbitt,dahan,bronkmann}.
%% It is quite possible that 
Possibly, a better understanding of the 
%understanding the somewhat elusive 
microscopic mechanism of the anomalous transport
%of the power law waiting time statistics 
%can be achieved by a combination of optical and transport measurements 
%on the same system, e.g. the fluorescence intermittency statistics 
%and the current noise.
can be achieved by establishing a connection between the 
statistics of fluorescence and of charge transmission in the same sample.
This could discriminate between transport due to the properties
of electron states in a single nanocrystal, and the 
collective transport phenomena.

%Continuous time random walks with power law WTDs (\ref{def-p}) 
%also arise when the system
%dynamics is determined by a broad distribution 
%of time scales, e.g. trap escape times 
%in amorphous solids yielding dispersive transport in
%photoconduction \cite{dispersive-transport}.
%Consider a simple trap model, 
%in which the escape time depends on energy $\epsilon$ via an activation exponent, 
%$\tau = \tau_0\, e^{\beta\epsilon}$. In such a system,  
%the density of states $\nu(\epsilon)=\nu_0 e^{-b\epsilon}$
%yields the WTD $p(\tau)$ of the form (\ref{def-p}) with 
%$\mu = b/\beta$ and $a = \nu_0\,\tau_0^{\mu}/\beta$.
%
%The latter mechanism, however, results in a current that 
%{\it grows} with time \cite{dispersive-transport}
%when a constant bias is applied. 
%However, in the case of a constant supply of carriers 
%the current {\it grows} with time \cite{street} at a constant bias.
%This is explained by gradually filling in of the traps thus smoothing the 
%potential for the carriers, effectively increasing the sample
%conductivity with time.
%At a {\it constant  bias}, however, such systems 
%are characterized by the {\it ohmic} conductivity that 
%The latter can slowly vary with time (bias stress) due to trapping instabilities
%\cite{trapping-instabilities}.

%%%%%%%%%%%%%%%%%%%%%%%%%%%%%%%%%%%%%%%%%%%%%%%%%%%%%%%%%%%%%%%%%%%%%%%%%%%%%%%%
\section{Conclusions}
\label{sec:conclusions}

\nin
This article presents a novel mechanism  
for a non--ohmic conductivity in a disordered system. 
In particular, we show that a non--stationary current 
response can arise in a stationary system with the L\'evy 
statistics of waiting times. 
The model agrees well with
%is found to be in a good agreement with 
the current and noise measurements
in arrays of coated semiconducting nanocrystals. 
The non-Poissonian character of the L\'evy process manifests itself
in the non-ohmic character of transport observed as the  
current transients, in memory effects, and in the colored noise.
Our results suggest that one needs to be careful in interpreting
conductivity in such systems\cite{sionnest-science} 
using simple ohmic models implying Poissonian statistics of transmission. 
We also demonstrate that the L\'evy model 
can help to investigate the system even 
without precise knowledge of microscopic transport mechanism, by linking the 
power law observed in the noise with that of current transient.

%Our results suggest that the notion of conductivity in QDAs 
%(used e.g. in Ref.~\onlinecite{sionnest-science})
%may be ill--defined since the latter implies Poissonian statistics in transmission.
%% We also demonstrate that noise can help to investigate the system even 
%% without precise knowledge of its microscopic transport mechanism.
%The microscopic origin of the power law statistics 
%% The latter is yet to be clarified by further studies.

\acknowledgments
This work was supported primarily by the MRSEC Program  
of the National Science Foundation under award number DMR 02-13282.
D.N. acknowledges support by NSF MRSEC grant DMR 02-13706. 
M.D. appreciates financial support from the Pappalardo Fellowship
and the ONR Young Investigator Award N00014-04-1-0489.

%%%%%%%%%%%%%%%%%%%%%%%%%%%%%%%%%%%%%%%%%%%%%%%%%%%%%%%%%%%%%%%%%%%%%%%%%%%%
\appendix

%%%%%%%%%%%%%%%%%%%%%%%%%%%%%%%%%%%%%%%%%%%%%%%%%%%%%%%%%%%%%%%%%%%%%%%%%%%%
\section{Current and noise in a single channel}
\label{sec:noise-calc}

\nin
%Below we show that the model with the WTD (\ref{def-p}) 
%yields a power law (\ref{I-exp}) in average current with $\alpha=1-\mu$.
Consider the current in a {single channel}
\be \label{def-I}
I(t) = \sum_{n=1}^{\infty} \delta(t-t_n) \,e^{-\lambda t} , 
%% \quad \lambda = T^{-1} \, . 
\ee
Here, instead of switching the current on and off at $t=0,T$ as in (\ref{def-I1}), 
we introduced a {\it soft cutoff} $\lambda = T^{-1}$. %time $\lambda^{-1}$ 
%corresponding to the measurement
%interval $T$. 
This cutoff helps to simplify calculations 
without qualitatively affecting the results.
The Fourier harmonic of (\ref{def-I}) is
\be \label{def-I-omega}
I_{\omega} = \int e^{-i\omega t} I(t)dt = 
\sum_{n=1}^{\infty} e^{-z t_n} \, ,
\ee
where $z = \lambda -i\omega$ and $t_n=\sum_{i=1}^n \tau_i\,$. 
Since waiting times $\tau_i$ are independent random variables
distributed according to $p(\tau)$, average current
is given by the geometric series
\be \label{I-ave-gen}
\la{I_{\omega}}\ra = {p_z\over 1 - p_z } \, ,
\ee
with $p_z$ the characteristic function 
\be \label{def-pz}
p_z \equiv \la e^{-z\tau}\ra = \int e^{-z\tau} p(\tau) d\tau
\, .
\ee
The correlator 
$\la I_{-\omega}I_{\omega}\ra = \sum_{n, n'=1}^{\infty} 
\la e^{-\bar z t_{n'} - z t_n} \ra$ 
can be evaluated as
\be \label{I-correlator}
\la I_{-\omega}I_{\omega}\ra 
= \sum_{n=1}^{\infty} \la e^{-2\lambda\tau_n} \ra
\lp 1 + \sum_{m=1}^{\infty} \la e^{- z\tau}\ra^m + {\rm c.c.} \rp.
\ee
The last formula is obtained by splitting the summation into parts with
$n=n'$ and $n>n'$ with $m=n-n'$.
The variance $\la\la I_{-\omega}I_{\omega}\ra\ra 
= \la I_{-\omega}I_{\omega}\ra - \la{I_{-\omega}}\ra \la{I_{\omega}}\ra $ 
is given by 
\be \label{I-variance-gen}
\la\la I_{-\omega}I_{\omega}\ra\ra  = 
\frac{p_{2\lambda}-p_z p_{\bar z}}{(1-p_{2\lambda})(1-p_z)(1-p_{\bar z})} \, .
\ee
where $p_{2\lambda}=p_{z=2\lambda}$.

The expressions for current average (\ref{I-ave-gen}) 
and variance (\ref{I-variance-gen}) are 
valid for any waiting time distribution.
Consider now the WTD of the form (\ref{def-p}). 
In this case the characteristic function is
\be
p_z = 1 - \frac{z^{\mu}}{A_{\mu}} \, ,
\quad A_{\mu} = \frac{\mu}{a\Gamma(1-\mu)} \, , 
%% \, , \quad |z|\tau_0 \ll 1 \, .
\ee
where $|z|\tau_0 \ll 1$, corresponding to the long time tail.
Eq.~(\ref{I-ave-gen}) then yields %the average current Fourier harmonic  
\be \label{I-ave}
\la{I_{\omega}}\ra = A_{\mu}\; (\lambda-i\omega)^{-\mu} ,
%% \quad A_{\mu} = \frac{\mu}{a\Gamma(1-\mu)} \, ,
\ee
%In this case the average current at $t\ll T$ is
resulting in the average current of the form (\ref{I-exp}):
\be \label{I-ave-t}
\la{I(t)}\ra = {\cal I}_0\; t^{-\alpha} \, , 
\quad \alpha = 1 - \mu \, , 
\quad {\cal I}_0 = \frac{\mu\sin \pi\mu}{\pi a} \, .
\ee
The Fourier harmonic variance,
obtained from Eq.~(\ref{I-variance-gen}), is of the form
%% yields for the variance
\be \label{I-variance1}
\la\la I_{-\omega}I_{\omega}\ra\ra = 
A_{\mu}^2 \;
\frac{z^{\mu}+\bar z^{\mu}-(2\lambda)^{\mu}-A_{\mu}^{-1}|z|^{2\mu}}
{(2\lambda)^{\mu} \, |z|^{2\mu}} \, .
\ee
We are interested in the noise spectrum on the time scale much greater
than the pulse width $\tau_0$: $\omega, \lambda \ll \tau_0^{-1}$.
Keeping the leading terms in Eq.~(\ref{I-variance1}),
we have
\be \label{I-variance2}
\la\la I_{-\omega}I_{\omega}\ra\ra = 
A_{\mu}^2 \;
\frac{z^{\mu}+\bar z^{\mu}-(2\lambda)^{\mu}}
{(2\lambda)^{\mu} \, |z|^{2\mu}} \, .
\ee
The limits of Eq.~(\ref{I-variance2}) are [$T = \lambda^{-1}$]:
\be \label{I-noise}
\la\la I_{-\omega}I_{\omega}\ra\ra = 
\cases{ \lp 2^{1-\mu}-1\rp  A_{\mu}^2 \; T^{2\mu} \, ,
& $\omega T \ll 1$ , \cr
2^{1-\mu} A_{\mu}^2 \cos\frac{\pi\mu}2 
\; T^{\mu}\omega^{-\mu} \, , 
 & $\omega T \gg 1$ .} 
\ee
%Thus for the low frequency noise spectrum we have 
%\be \label{I-noise-low-freq}
%\la\la I_{-\omega}I_{\omega}\ra\ra = 
%\la I_{\omega}\ra^2 \cdot
%\lp  \frac{(1+i\omega T)^{\mu} + (1-i\omega T)^{\mu}}{2^{\mu}}  - 1\rp \, . 
%\ee
% 
%This corresponds to a large (non-Poissonian) noise
%proportional to average current.
%
The result (\ref{I-ave-t}) corresponds to the net transmitted charge 
\be \label{Q-T}
\la Q\ra = A_{\mu} T^{\mu} \, . 
\ee
For the exponent $\mu < 1$, all the moments of (\ref{def-p}), 
including the mean and the variance, diverge, 
and thus the central limit theorem does not hold.
Instead of $\la Q\ra\propto T$ expected for a stationary random process,
here we have a power law.
Moreover, the distribution of $Q(T)$ is extremely broad,
with dispersion proportional to the net charge:
\be \label{disp-Q}
{\rm var}\, Q = \lp \la Q^2\ra-\la Q\ra^2\rp^{1/2}
= \lp 2^{1-\mu}-1\rp^{1/2} \; \la Q\ra \, .
\ee
Hence the ratio 
${\rm var}\, Q/\la Q\ra$ does not decrease with time $T$, 
violating the central limit theorem.

The large frequency asymptotic behavior of the current and noise is given 
by the same characteristic power law. 
According to Eq.~(\ref{I-noise}), the noise-to-current ratio
is controlled by the net transmitted charge (\ref{Q-T}) through the channel:
\be \label{r}
r \equiv \frac{\la\la I_{-\omega}I_{\omega}\ra\ra}{|\la I_{\omega} \ra|}
= c \, Q(T) \, , \quad c= 2^{1-\mu} \cos{\pi\mu\over 2} \, .
\ee

It is instructive to compare our results for the WTD (\ref{def-p})
with those derived for the Poissonian statistics.
In the Poissonian case,  
$p(\tau) = \bar\tau^{-1} e^{-\tau/\bar\tau}$,
$p_z = (1+ z\bar\tau)^{-1}$.
Eq.~(\ref{I-ave-gen}) yields the average current
$\la I_{\omega}\ra = \lb(\lambda-i\omega)\bar\tau\rb^{-1}$,
corresponding to $\la I(t)\ra = 1/\bar\tau$ for  $t\ll T= \lambda^{-1}$.
At the same time, 
Eq.~(\ref{I-variance-gen}) yields the white-noise spectrum
$\la\la I_{-\omega}I_{\omega}\ra\ra = T/2\bar\tau
= \frac12 \la I_{\omega=0}\ra$, in agreement with the central limit theorem.

%%%%%%%%%%%%%%%%%%%%%%%%%%%%%%%%%%%%%%%%%%%%%%%%%%%%%%%%%%%%%%%%%%%%%%%%%%%%%%%%%

\end{document}